\documentclass{appolb}
\usepackage{epsfig}
\usepackage{graphicx}
\usepackage{hyperref}
\usepackage{subfigure}
\usepackage{xspace}
\usepackage{color}
\usepackage{amsmath}
\usepackage{amssymb}
\graphicspath{
{./figures/}
}

\newcommand{\Dslash}{\ensuremath{D\hspace{-1.5ex} /}}

\def\roughly#1{\mathrel{\raise.3ex\hbox{$#1$\kern-.75em%
\lower1ex\hbox{$\sim$}}}}

\newcommand{\vev}[1]{\ensuremath{\left\langle #1 \right\rangle}}
\newcommand{\einh}[1]{\ensuremath{\,\text{#1}}}
\newcommand{\MeV}{\einh{MeV}}

\def\Eq#1{Eq.~(\ref{#1})}
\def\Fig#1{Fig.~\ref{#1}}

\newcommand{\ua}{\ensuremath{U(1)_A}}
\newcommand{\Phibar}{\ensuremath{\bar{\Phi}}}
\newcommand{\LPQM}{\ensuremath{\mathcal{L}_{\textrm{PQM}}}\xspace}

\def\Eq#1{Eq.~(\ref{#1})}
\def\Fig#1{Fig.~\ref{#1}}

\newcommand{\twofigs}{0.48\linewidth}

\newcommand{\Tl}{\ensuremath{T_\chi}}

\newcommand{\Tchi}{\ensuremath{T_\chi}}

\newcommand{\coloronl}{}
\newcommand{\muT}{\left(\frac{\mu}{T}\right)}

\begin{document}
\title{QCD Thermodynamics: Confronting the Polyakov-Quark-Meson Model with Lattice QCD
\thanks{Presented at 'Three Days of Strong Interactions', 
July 9 - 11, 2009, Wroclaw, Poland}%
}
\author{J. Wambach
\address{Institut f\"ur Kernphysik, TU Darmstadt, D-64289 Darmstadt, Germany}
\and
B.-J. Schaefer
\address{Institut f\"ur Physik, Karl-Franzens-Universit\"at, A-8010
  Graz, Austria}
\and
M. Wagner
\address{Institut f\"ur Kernphysik, TU Darmstadt, D-64289 Darmstadt, Germany and\\
        ExtreMe Matter Institute EMMI,  GSI Helmholtzzentrum f\"{u}r Schwerionenforschung GmbH, D-64291 Darmstadt, Germany}
}

\maketitle
\begin{abstract}
  NJL-type effective models represent a low-energy realization of QCD
  and incorporate pertinent aspects such as chiral symmetry and its
  spontaneous breaking, the center symmetry in the heavy-quark limit
  as well as the axial anomaly. One such model, the
  Polyakov-quark-meson model for three light quark flavors, is
  introduced in order to study the phase structure of
  strongly-interacting matter.
  With recent high-statistics lattice QCD simulations of the
  finite-temperature equation of state, a detailed comparison with
  model results becomes accessible. 
  Such comparisons allow to estimate volume and truncation effects of
  quantities, obtained on the lattice and provide possible lattice
  extrapolation procedures to finite chemical potential which are
  important to locate a critical endpoint in the QCD phase diagram.
\end{abstract}

  
\section{Introduction}

A detailed theoretical understanding of strongly-interacting matter
under extreme conditions is mandatory for various heavy-ion research
programs. The search for possible (tri)critical endpoints in the phase
diagram is a major focus of the CBM experiment at the future FAIR
facility.

Different regimes of the QCD phase diagram can be explored by
employing various theoretical methods. Lattice QCD simulations are
applicable at zero or imaginary chemical potentials. 
Most simulations for $(2+1)$-flavor QCD agree
that chiral symmetry is restored by a smooth crossover
transition at $\mu=0$~\cite{Aoki:2006we} but there is still an ongoing discussion
concerning the (pseudo)critical temperatures~\cite{Cheng:2006qk}. 
At finite real
chemical potentials the fermion sign problem remains a considerable
obstacle. Several extrapolation techniques to finite chemical
potentials such as the reweighting method, imaginary chemical
potential or a Taylor expansion around vanishing chemical potentials
have been proposed (for an overview see~\cite{Philipsen:2005mj}). 
Effective models, such as the Polyakov-loop NJL
(PNJL) or the Polyakov-loop quark-meson (PQM) model~\cite{Schaefer:2007pw}, incorporate
important fundamental symmetries and the symmetry breaking pattern
of the underlying QCD but do not suffer from the sign problem, large
quark masses or finite volume restrictions. Furthermore, these type of
models can be used to test certain lattice extrapolation techniques
to finite $\mu$. For this purpose the reproduction of the lattice data
for vanishing $\mu$ is a basic prerequisite.

In this talk we compare the bulk thermodynamics of several chiral
$N_f=2+1$ quark flavor PQM models with recent $N_\tau=8$ lattice data
of the HotQCD collaboration~\cite{Bazavov:2009zn}. The larger quark
masses used on the lattice are explicitly considered in the model
comparison. Based on a novel differentiation
technique, higher derivatives of the thermodynamic potential can be
calculated very precisely \cite{ADTAylor}. This method allows to
investigate convergence properties of the Taylor expansion method used
on the lattice.

\section{Polyakov-Quark-Meson Model}
\label{sec:pqm}

The Polyakov-quark-meson (PQM) model for three quark flavors is based
on the linear $\sigma$-model with quarks~\cite{Schaefer:2008hk}
 and incorporates in addition the Polyakov loop
field $\Phi(\vec x)$. The Polyakov loop is the thermal expectation
value of a color traced Wilson loop in the temporal direction. In the
heavy-quark limit $\Phi$ serves as an order parameter for the
confinement/deconfinement transition. In the deconfined
high-temperature phase the center symmetry $Z(3)$ of QCD is
spontaneously broken and as a consequence $\Phi$ is finite. However,
with dynamical quarks the center symmetry is always broken and $\Phi$
as an order parameter becomes questionable.

The PQM Lagrangian consists of a quark-meson contribution and a
Polyakov-loop potential $\mathcal{U} (\Phi, \Phibar)$, which depends
on $\Phi$ and its hermitian conjugate $\Phibar$. The uniform temporal
background gauge field is coupled to the quarks by replacing the
standard derivative $\partial_\mu$ in the quark contribution by a
covariant derivative
$D_\mu = \partial_\mu - i A_\mu, \; A_\mu = \delta_{\mu 0}A^0$ where
$A_\mu\equiv g_s A_\mu^a \lambda^a/{2}$. This leads to the Lagrangian
\begin{equation}
\label{eq:lpqm}
\LPQM = \bar{q}\left(i \Dslash - g \phi_5 \right) q +
\mathcal{L}_m  
-\mathcal{U} (\Phi[A], \Phibar[A])\ ,
\end{equation}
where $q$ denotes the quark field. The interaction between the quarks
and the meson nonets is implemented by a flavor-blind Yukawa coupling
$g$ and the meson matrix
$\phi_5 = \sum_{a=0}^8(\lambda_a/2)\left( \sigma_a + i \gamma_5
  \pi_a\right)$ where the nine scalar mesons fields are labeled by
$\sigma_a$ and accordingly the nine pseudoscalar fields by $\pi_a$.

The remaining, purely mesonic contribution reads
\begin{eqnarray}
\label{eq:mesonL}
  \mathcal{L}_m &=& \Tr \left( \partial_\mu \phi^\dagger \partial^\mu
    \phi \right)
  - m^2 \Tr ( \phi^\dagger \phi) -\lambda_1 \left[\Tr (\phi^\dagger
    \phi)\right]^2 - \lambda_2 \Tr\left(\phi^\dagger \phi\right)^2
\nonumber \\
  &&    +c   \left(\det (\phi) + \det (\phi^\dagger) \right)+ \Tr\left[H(\phi+\phi^\dagger)\right]\ ,
\end{eqnarray}
with the fields
$\phi\equiv \sum_a(\lambda_a/2) \left(\sigma_a + i \pi_a\right)$.
Chiral symmetry is explicitly broken by the last term in
\Eq{eq:mesonL} and the $\ua$-symmetry by the 't Hooft determinant term
with constant strength $c$.

For the effective Polyakov loop potential $\mathcal{U}$ which is
constructed in terms of $\Phi$ and $\Phibar$, several implementations
are available. The simplest choice is based on a Ginzburg-Landau
ansatz \cite{Ratti:2005jh}:
\begin{equation}
\label{eq:upoly}
  \frac{\mathcal{U}_{\rm{poly}}}{T^{4}}= -\frac{b_2}{4}
\left(|\Phi |^2+|\bar\Phi |^2 \right) 
-\frac{b_3}{6}(\Phi^3+\Phibar^3)+\frac{b_4}{16}
\left(|\Phi|^2+|\Phibar|^2\right)^2\ .
\end{equation}
The cubic $\Phi$ terms are required to break the $U(1)$ symmetry of
the remaining terms down to the center symmetry, $Z(3)$. The potential
parameters are adjusted to the pure gauge lattice data such that the
equation of state and the Polyakov loop expectation values at finite
temperature are reproduced. An improved version \cite{Roessner:2006xn}
based on the $SU(3)$ Haar measure results in
\begin{equation}
\label{eq:ulog}
\frac{\mathcal{U}_{\rm{log}}}{T^{4}}= -\frac{1}{2}a(T) \Phibar \Phi \\
+ b(T) \ln \left[1-6 \Phibar\Phi + 4\left(\Phi^{3}+\Phibar^{3}\right)
  - 3 \left(\Phibar \Phi\right)^{2}\right]\ .
\end{equation}
The parameters are again fitted to the pure gauge lattice data. The
logarithmic form constrains $\Phi$ and $\bar\Phi$ to values smaller
than one.
Another choice invented by Fukushima~\cite{Fukushima:2008wg} is
\begin{equation}
\label{eq:ufuku}
\frac{\mathcal{U}_{\rm{Fuku}}}{T^4} = -\frac{b}{T^3}  \left[54 e^{-a/T} \Phi \Phibar  \right.\\
\left.+ \ln \left(1 - 6 \Phi \Phibar - 3 (\Phi \Phibar)^2 + 4 (\Phi^3
    + \Phibar^3)\right)\right]
\end{equation}
with only two parameters $a$ and $b$. The parameters also result in a
first-order transition at $T_0 \sim 270 \MeV$ in the pure gauge sector
but are not fitted to lattice data. This potential excludes
contributions of the unconfined transverse gluons to the equation of
state which are relevant at high temperatures~\cite{Fukushima:2008wg}.
This is important for the comparison to lattice data at higher
temperatures.

\subsection{Thermodynamic Potential}
\label{sec:pot}

For three quark flavors the grand potential $\Omega$ is a function of
the temperature and in general three quark chemical potentials, one
for each flavor. Here we focus on a uniform quark chemical potential
$\mu\equiv\mu_q=\mu_B/3$. Since we consider the isospin-symmetric case
with $m_l \equiv m_u=m_d$, only two order parameters, the non-strange
$\sigma_x$ and strange $\sigma_y$ emerge~\cite{Schaefer:2008hk}.
The thermodynamic potential in mean-field
consists of three different contributions~\cite{Schaefer:2009ui}: the mesonic part
$U\left(\sigma_{x},\sigma_{y}\right)$, a quark part
$\Omega_{\bar{q}{q}}$ and the Polyakov loop potential
\begin{equation}
  \label{eq:grandpot}
  \Omega = U \left(\sigma_{x},\sigma_{y}\right) +
  \Omega_{\bar{q}{q}} \left(\sigma_{x},\sigma_{y}, \Phi,\Phibar \right) +
  \mathcal{U}\left(\Phi,\Phibar\right)\ . 
\end{equation}
The mesonic contribution has six parameters which are fitted to the
vacuum. For example, the Yukawa coupling $g$ is fixed to reproduce a
light constituent quark mass of $m_{l} \approx 300$ MeV. This then
yields a strange constituent quark mass of $m_s \approx 433$ MeV.

The temperature- and quark chemical potential dependence of the four
order parameters for the chiral and confinement/deconfinement
transition are determined as solutions of the corresponding gap
equations. These coupled equations are obtained by minimizing the
grand potential, \Eq{eq:grandpot}, with respect to the four constant
mean-fields $\vev \sigma_x$, $\vev \sigma_y$, $\vev\Phi$ and $\vev{\bar\Phi}$:
\begin{equation}
  \label{eq:pqmeom}
  \left.\frac{ \partial \Omega}{\partial 
      \sigma_x} = \frac{ \partial \Omega}{\partial \sigma_y}  = \frac{
      \partial \Omega}{\partial \Phi}  =\frac{ \partial
      \Omega}{\partial \Phibar} 
  \right |_{\rm{min}} = 0\ ,
\end{equation}
where
$\rm{min}=\left\{\sigma_x=\vev{\sigma_x}, \sigma_y=\vev{\sigma_y},
  \Phi=\vev{\Phi}, \bar\Phi=\vev{\bar\Phi} \right\}$ labels the global
minimum.

\section{Lattice comparison}

In order to compare our model results with recent HotQCD lattice
findings~\cite{Bazavov:2009zn} we have chosen a model parameter setup
where the chiral and deconfinement transition temperature coincide at
$\mu=0$. This is the case for $m_\sigma=600\MeV$ and $T_0=270\MeV$ for
all used Polyakov-loop potentials~\cite{Schaefer:2009ui}. A coincidence of
both transitions around $T_\chi \sim 185-195 \MeV$ is also observed in the
corresponding lattice simulations~~\cite{Bazavov:2009zn}. However, on the
lattice a ratio of the physical strange quark mass to the light one of
10 has been used which yields finally too heavy light-quark
masses. For a proper comparison we have therefore adjusted
the pion and kaon masses in the model calculations accordingly and use
also $m_K= 503\MeV$ and $m_\pi=220 \MeV$.
Of course, heavier meson masses yield also slightly heavier
constituent quark masses
\begin{equation*}
m_l \approx 322 \MeV \quad \text{and} \quad m_s \approx 438\MeV \ .
\end{equation*}
As a consequence, higher transition temperatures, in particular for
the chiral transition in the light sector, are found. But both
transitions, the non-strange chiral and the deconfinement one, still
coincide. The strange quark sector is almost unaffected~\cite{Schaefer:2009ui}.


\begin{figure}[tbp!]
\centering
\subfigure[$\ $ Normalized pressure $p/p_{SB}$] {\label{sfig:pressure}
  \includegraphics[width=\twofigs]{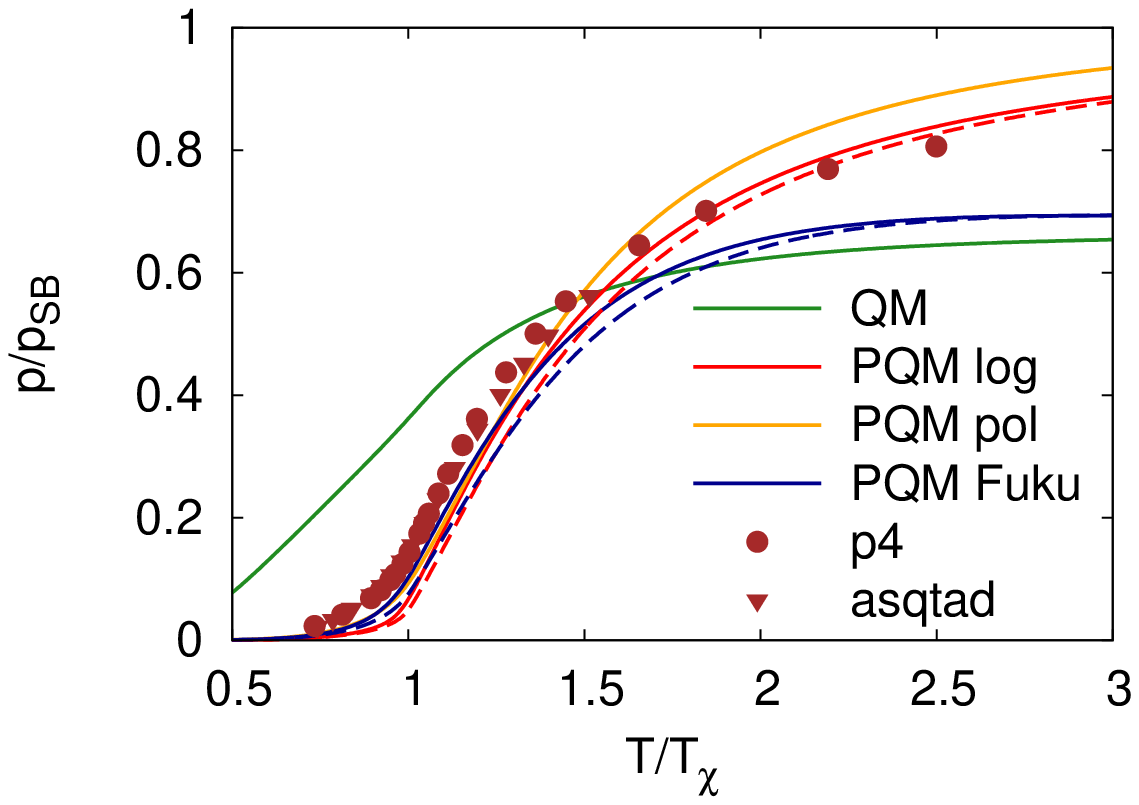}}
\subfigure[$\ $ Interaction measure $\Delta/T^4$]
{\label{sfig:e3p}
  \includegraphics[width=\twofigs]{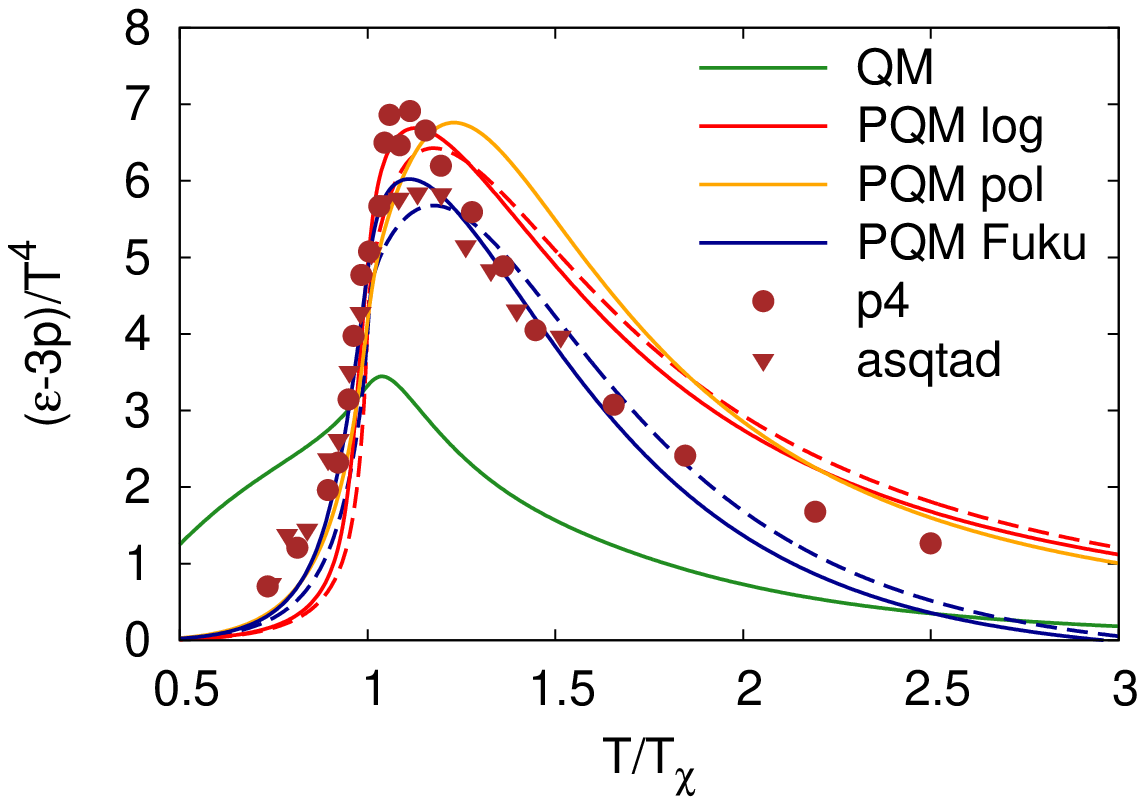}}
\caption{\label{fig:pqmeos}\coloronl The normalized pressure (left
  panel) and the interaction measure (right panel) as a function of
  temperature. The model calculations (PQM model with various Polyakov-loop potentials and the QM model) are compared to lattice data
  ($N_\tau=8$, p4 and asqtad actions) from ~\cite{Bazavov:2009zn}.
  The solid lines correspond to larger pion and kaon masses as used in the
  lattice simulations while the dashed lines denote the results for physical masses.}
\end{figure}

In the model calculations all thermodynamic quantities are extracted
from the grand potential. The pressure
\begin{equation}
p(T,\mu) = - \Omega\left(T,\mu \right)
\end{equation}
is directly related to the thermodynamic potential with a suitable
normalization, $p(0,0) = 0$. In \Fig{fig:pqmeos} the pressure,
normalized to the Stefan-Boltzmann (SB) value of the PQM model, is
compared to the lattice data and a quark-meson (QM) model calculation.
As expected, the QM model~\cite{Schaefer:2008hk} fails in describing
the lattice data, while the PQM model results are in better agreement
with the data. The best agreement is achieved with Fukushima's
potential where the different treatment of the transverse gluons at
higher temperatures is apparent. Around the transition the model
versions with the polynomial or logarithmic Polyakov loop
potential are closer to the lattice data. For physical meson masses
(dashed lines) the pressure is smaller.

The interaction measure, $\Delta = e-3p$, indicates the breaking of
scale invariance and is given as a temperature derivative of the grand
potential
\begin{equation}
\label{eq:e3p}
\frac{ \Delta}{T^4} = T \frac{\partial}{\partial
  T}{\left(\frac{p}{T^4}\right)} = -T \frac{\partial}{\partial
  T}\left( \frac{\Omega}{T^4} \right) \ .
\end{equation}
In lattice simulations, this quantity can be obtained directly from
the trace of the energy-momentum tensor
\begin{equation} 
  \frac{\Theta^{\mu}_{\mu}}{T^4}= \frac{\epsilon - 3p}{T^4} = \frac{
    \Delta}{T^4}\ .
\end{equation}
It is more sensitive to finite volume and discretization effects than,
e.g., the pressure (cf.~\cite{Bazavov:2009zn}) which is displayed in
\Fig{sfig:e3p}. In the chirally broken phase, the lattice data are
close to all model curves. Around the transition $T \approx T_\chi$
Fukushima's potential is closest to the asqtad-action data, while the
remaining two model curves are in better agreement with the p4-action
data. In general, the peak height decreases on larger lattices, in
particular for the p4-action while the asqtad-action 
shows a weaker $N_\tau$ dependence \cite{Bazavov:2009zn}. In contrast
to the pressure, the logarithmic and polynomial Polyakov model version
do not describe the lattice data in the symmetric phase. However,
Fukushima's ansatz approaches the data at least up to $T\sim1.5 \Tl$.

\begin{figure*}[tbp]
  \centering \subfigure[$\ $Energy density
  $\epsilon/T^4$]{\label{sfig:epsilon}\includegraphics[width=\twofigs]{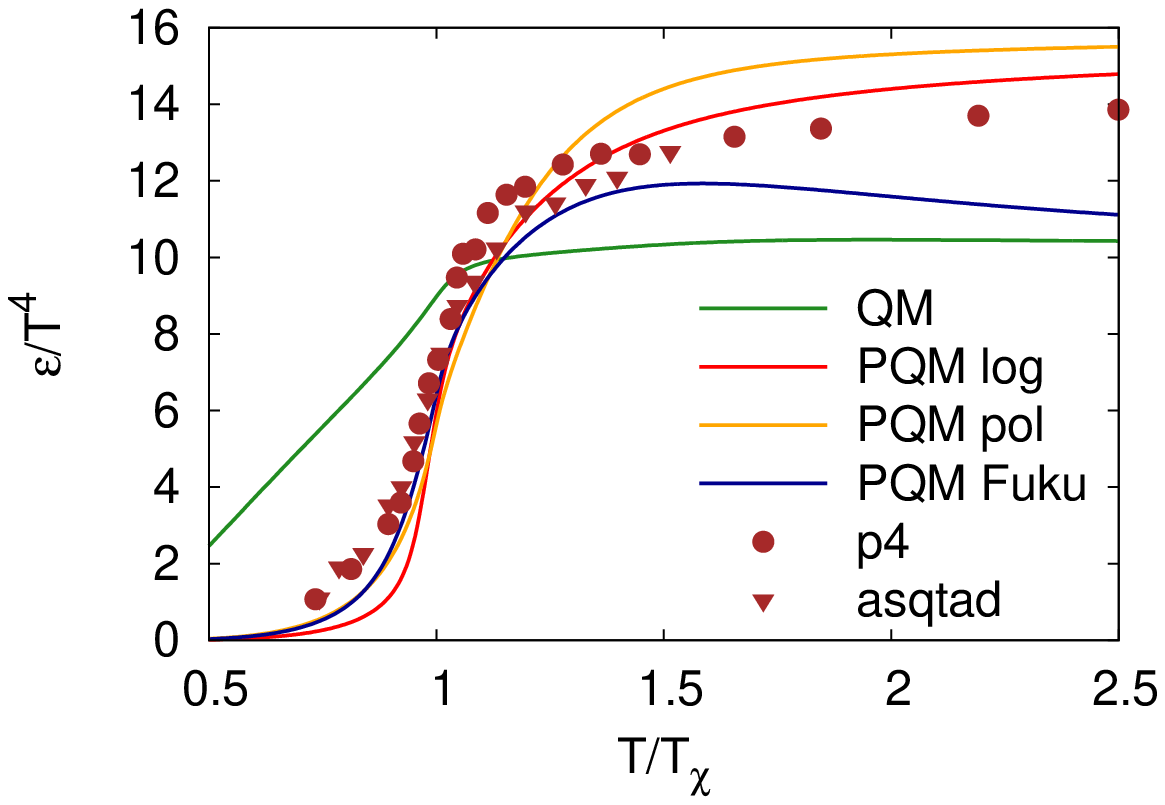}} 
  \subfigure[$\ $Entropy density
  $s/T^3$]{\label{sfig:entropy}\includegraphics[width=\twofigs]{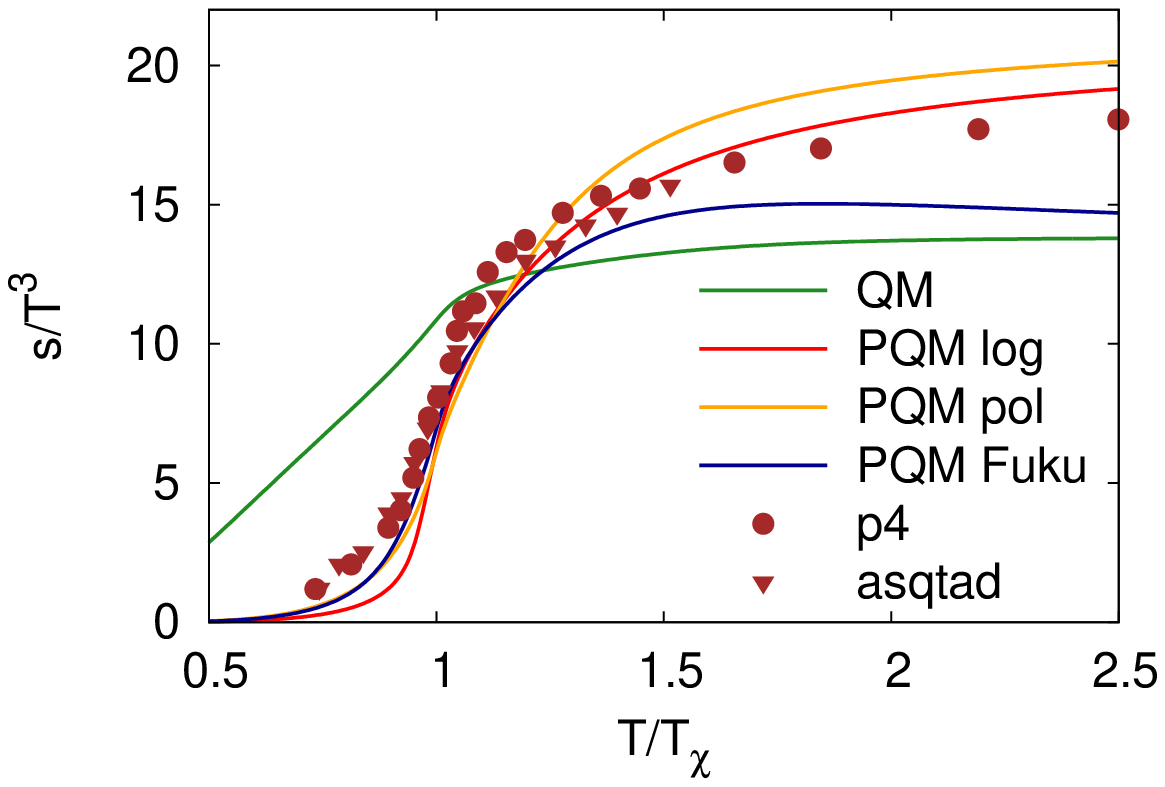}}
  \caption{\label{fig:edensity}\coloronl The energy (left panel) and
    entropy density (right panel) similar to Fig.~\ref{fig:pqmeos}.}
\end{figure*}

In addition we also show the energy density $\epsilon$ and the entropy
density $s$, which are defined as
\begin{equation}
\label{eq:energydensity}
\epsilon = -p + T s \quad , \quad s = - \frac{\partial
  \Omega}{\partial T} \ ,
\end{equation}
in \Fig{fig:edensity} as a function of temperature. Similar to the
pressure, Fukushima's potential model version comes closest to the
lattice data for $T \lesssim 1.5 \Tchi$.

\section{Convergence of the Taylor expansion at finite $\mu$}

At finite chemical potential the spectrum of the Dirac operator
becomes complex making direct Monte Carlo simulations impossible.
Several methods have been developed to circumvent this problem 
and to access at least a region of small chemical
potentials in the phase diagram, see e.g.~
\cite{Philipsen:2005mj}.

One such approach is based on the Taylor expansion of thermodynamic
quantities in powers of $(\mu/T)$ around $\mu=0$. For the pressure it
reads
\begin{equation}
  \frac{p(\mu/T)}{T^4} = \sum_{n=0}^\infty c_n(T) \muT^n
\end{equation}
with the coefficients
\begin{equation}
  c_n(T) = \left.\frac{1}{n!} \frac{\partial^n\left(p(T,\mu)
        /T^4 \right)}{\partial \left(\mu/T\right)^n}
  \right|_{\mu=0}\ . 
\end{equation}
Note that only even coefficients contribute since the QCD partition
function is CP-symmetric, i.e., $Z(\mu) = Z(-\mu)$. The coefficients
have been calculated up to the eighth order by different lattice
groups, for $N_f=2$ see
e.g.~\cite{Allton2005gk} 
and for $N_f=2+1$
quark flavors, see e.g.~\cite{Miao:2008sz}. However, higher orders
still suffer from large errors. 

\begin{figure*}[tb]
\centering
\includegraphics[width=0.9\linewidth]{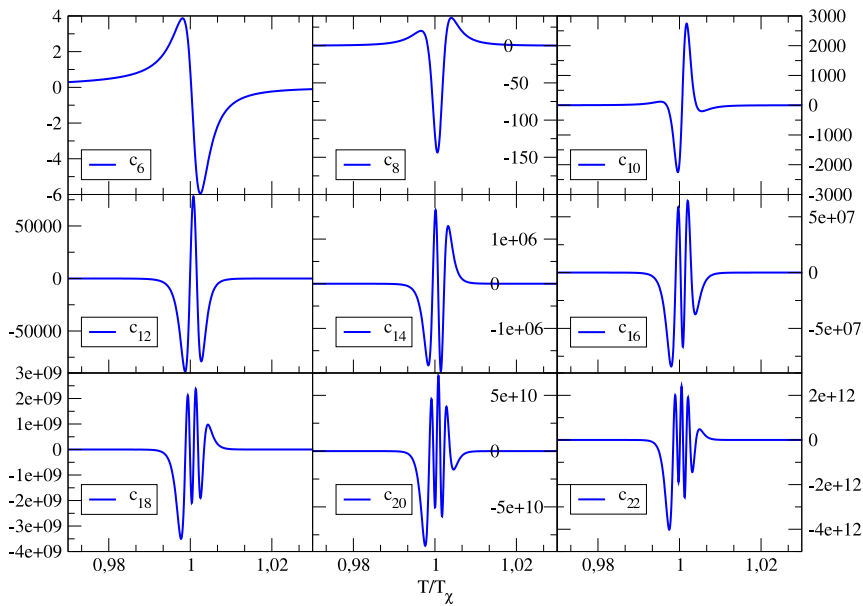} 
\caption{\label{fig:pqmtaylors2} The Taylor coefficients $c_6$ to
  $c_{22}$ obtained in the PQM model with the logarithmic Polyakov
  loop potential ($T_0=270\MeV, m_\sigma=600\MeV$).}
\end{figure*}

The application of the Taylor expansion method to the PQM model
provides an opportunity to investigate and test the finite $\mu$
extrapolation since a direct evaluation of the thermodynamic potential
in the model is possible. For example, the location of the critical
endpoint (CEP) is exactly known in the model calculation. This enables
us to distinguish between divergences related to the CEP and to the
breakdown of the expansion.

However, a further difficulty emerges in the model analysis: in
principle, the grand potential is known analytically but an analytic
evaluation of the Taylor coefficients fails since the implicit
temperature and chemical potential dependence of the order parameters
is only known numerically via the equations of motion \Eq{eq:pqmeom}.

Standard methods for an evaluation of numerical derivatives are
hampered by increasing errors, which become dominant in particular for
higher derivative orders. In order to circumvent this caveat we have
developed a novel numerical technique which is based on algorithmic
differentiation (AD). With this method the evaluation of higher
derivatives becomes feasible to extremely high precision. In fact,
it is essentially limited only by machine precision. Details of this
method can be found in Ref.~\cite{ADTAylor}.
 
With this method we could obtain the Taylor coefficients up to
$22^\textrm{nd}$ order for the PQM model with the logarithmic Polyakov
loop potential as shown in \Fig{fig:pqmtaylors2}. The coefficients are
very small far away from the transition temperature and they start to
oscillate with increasing amplitude within a narrow temperature range
around the transition temperature, i.e.~$0.95 \Tchi < T < 1.05 \Tchi$.
This signals a complex $\mu$ singularity of the thermodynamic
potential close to the real axis. In the chiral limit this singularity
would be exactly on the real axis as a reflection of a real phase
transition.

By means of the Taylor expansion method it is further possible to
study several thermodynamic quantities at small $\mu$. One example
is the pressure difference
\begin{equation}
  \frac{\Delta p(T,\mu)}{T^4} = \frac{p(T,\mu)-p(0,0)}{T^4} =
  \sum_{n=2,4,\ldots} c_n(T) \left(\frac{\mu}{T}\right)^n\ . 
\end{equation}

\begin{figure*}[tb]
  \centering \subfigure[$\muT=0.8$]{
    \includegraphics[width=\twofigs]{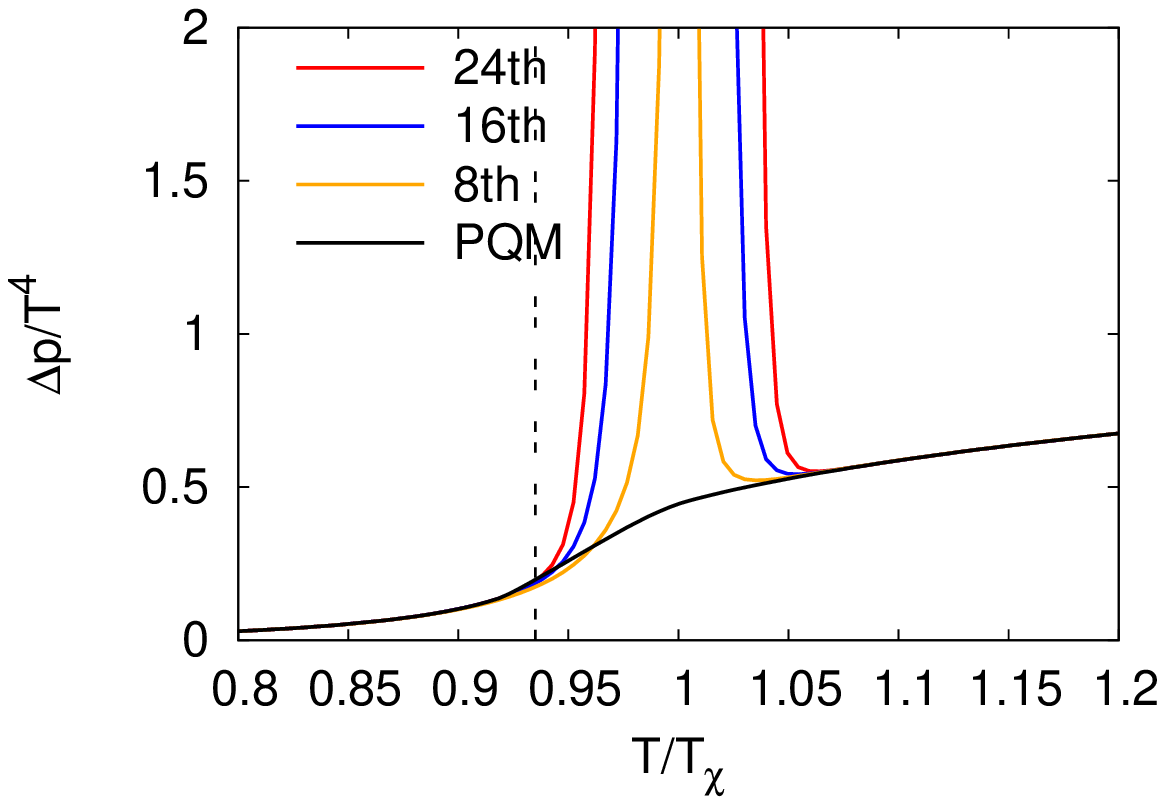}} 
  \subfigure[$\muT=1.0$]{
  \includegraphics[width=\twofigs]{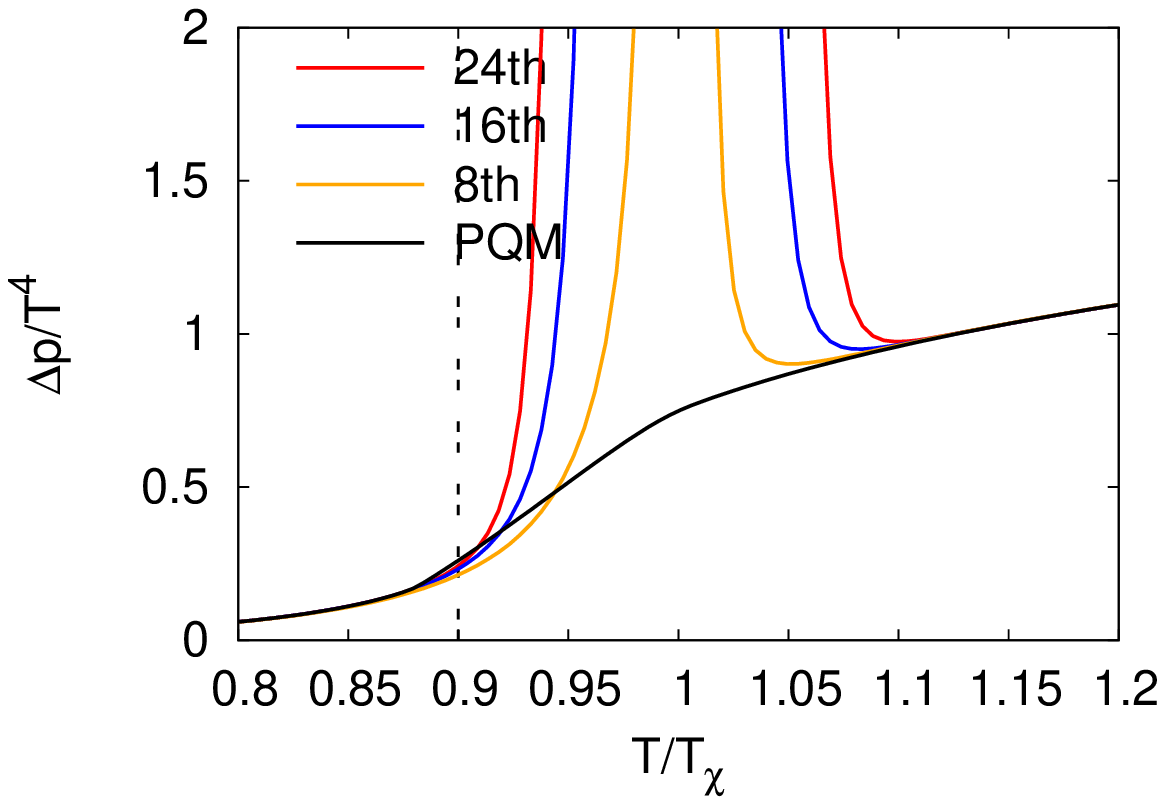}}
\caption{\label{fig:taylorpressure}The pressure difference
  $\Delta p / T^4$ for different ratios of $\mu/T$ and orders of the
  Taylor expansion. The black line, labeled with 'PQM', is the model
  calculation. The vertical dashed line indicates the radius of
  convergence.}
\end{figure*}

In \Fig{fig:taylorpressure} we show $\Delta p$ for $\mu/T=0.8$ and
$\mu/T=1.0$ for various expansion orders in comparison with the model
result. The ratios $\mu/T$ are slightly below and above the critical
value $\mu_c/T_c \sim 0.9$ at the CEP. Although the transition is
still a crossover at $\mu/T=0.8$, a divergence is observed in the
Taylor expansion. The divergence becomes more prominent with
increasing orders. However, a direct model evaluation of the pressure,
labeled as 'PQM' in the figure, shows a smooth behavior. The
divergence in the coefficients is clearly a signal for the breakdown
of the Taylor expansion, which is also related to the oscillations in
the coefficients. This occurs already for $\mu/T<1$ and indicates the
importance of the higher order coefficients. For temperatures just
below the breakdown at $T\sim0.95 \Tl$ the higher orders improve the
observed agreement with the 'PQM' curve. A similar behavior is also
seen for the $\mu/T=1$ case. At higher temperatures $T \gg T_\chi$ the
coefficients become small again and do not longer contribute. As a
consequence the Taylor expansion can again reproduce the PQM result.

For a deeper understanding of the breakdown of the Taylor expansion it
is instructive to investigate its convergence radius. It can be
obtained from the definition
\begin{equation}
  r = \lim_{n \rightarrow \infty} r_{2n} = \lim_{n \rightarrow \infty}
  \left|\frac{c_{2n}}{c_{2n+2}} \right|^{1/2}\ . 
\label{eq:convr}
\end{equation}
It is not yet known how well the radius $r$ is estimated by $r_n$ for
a finite value $n$. The estimated $r_{24}$ is indicated by the dashed
vertical line in \Fig{fig:taylorpressure}. It coincides with the
occurrence of the divergence observed in $\Delta p$ at $24$-th order.
From the different $r_n$ we expect that the true convergence radius is
smaller than $r_{24}$ for the given $\mu/T$ ratios. For more details
see ~\cite{Schaefer:2009st}. A detailed study concerning the prospects
of locating the CEP within the Taylor expansion will be given
in~\cite{Taylorcoeff}.

\section{Summary}

In this talk we have presented results for the bulk thermodynamics of
a ($2+1$)-flavor PQM-model with three different effective Polyakov
loop potentials. The model parameters are adjusted at vanishing
chemical potential to produce a coincidence of the chiral and
deconfinement transition for all three Polyakov loop potentials. For a
better comparison with the recent HotQCD lattice data we further tuned
the pion and Kaon masses to these values which are used in the lattice
simulations. All three PQM model versions yield a good reproduction of
the lattice data for temperatures $T<1.5\Tchi$. Furthermore, the
Taylor expansion of the pressure difference for finite chemical
potentials is also investigated. For this purpose, a novel algorithmic
differentiation technique has been developed. The Taylor coefficients
up to $24^\textrm{th}$ order could be obtained in the PQM model for
the first time. The knowledge of these higher Taylor coefficients
allows for a systematic study of convergence properties of the
expansion.

\subsubsection*{Acknowledgments}
The work of MW was supported by the Alliance Program of the Helmholtz
Association (HA216/ EMMI) and BMBF grants 06DA123 and 06DA9047I. JW
was supported in part by the Helmholtz International Center for FAIR. We further acknowledge the support of the European Community-Research
Infrastructure Integrating Activity
Study of Strongly Interacting Matter under the Seventh Framework Programme of EU.



\end{document}